\documentclass[3p,times,procedia]{elsarticle}
\usepackage{color}
\usepackage{xcolor}

\usepackage{ecrc}
\usepackage[bookmarks=false]{hyperref}
    \hypersetup{colorlinks,
      linkcolor=blue,
      citecolor=blue,
      urlcolor=blue}

\volume{00}

\firstpage{1}

\journalname{Procedia Computer Science}

\runauth{Author name}

\jid{procs}

\usepackage{amssymb}

\usepackage[figuresright]{rotating}

\begin{document}
\begin{frontmatter}



\dochead{24th International Conference on Knowledge-Based and Intelligent Information \& Engineering Systems}%

\title{Getting Insights from a Large Corpus of Scientific Papers on Specialisted Comprehensive Topics - the Case of COVID-19 Epidemia}
\title{Getting Insights from a Large Corpus of Scientific Papers on Specialisted Comprehensive Topics - the Case of COVID-19 }


\author[a]{Bernard Dousset} 
\author[b]{Josiane Mothe\corref{cor1}}

\address[a]{ IRIT, UMR5505, CNRS \& Univ. Toulouse, France}
\address[b]{ IRIT, UMR5505, CNRS \& INSPEE UT2J, Univ. Toulouse, France}

\begin{abstract}
COVID-19 is one of the most important topic these days, specifically on search engines and news. While fake news are easily shared, scientific papers are reliable sources where information can be extracted. With about 24,000 scientific publications on COVID-19 and related research on PUBMED, automatic computer-assisted analysis is required. In this paper, we develop two methodologies to get insights on specific sub-topics of interest and  latest research sub-topics. They rely on natural language processing and graph-based visualizations. We   run these methodologies on two cases: the virus origin and the uses of existing drugs.

%
\end{abstract}

\begin{keyword}
Topic analysis; Automatic mining of scientific publication; COVID-19; automatic keyphrase extraction;




\end{keyword}
\cortext[cor1]{Corresponding author. Tel.: +33-6-86724870  }
\end{frontmatter}

\email{Josiane.Mothe@irit.fr}



\section{Introduction}
\label{sec:introduction}

COVID-19 is one of the most important  topic these days, specifically on search engines and news. It is a worldly shared topic of interest. 
While news are looped on TV, a very few specialists know deeply on COVID-19.  On the Internet, a lot of fake news started to circulate and spread as fast as the virus it-self. In such situation, citizens and decision makers need reliable sources of information. Scientific papers are certainly such reliable sources that could be used for helping citizens knowing more about it and being informed on both a reliable and accurate way. 

Not only scientific sources can be use to try to answer some specific questions that arise but it is also a way to know  the main researchers, groups or institutes that work on a specific sub-topic, what the collaborations are, ... Both in- deep analyses and overviews on a large quantity of research papers can help decision makers or even citizens to be better educated on the state of the art. For people, it can help move aside fake news. It can also help new comers in the COVID-19 research field  providing  overviews of sub-topics first (main publication venues, main authors, ...

Indeed the research in the domain is quite huge specifically if we consider other forms of COVIDs, Severe Acute Respiratory Syndrome, and Middle East Respiratory Syndrome. For example, the recently released collection named COVID-19 Open Research Dataset~\footnote{\url{https://pages.semanticscholar.org/coronavirus-research}} consists of more than 40,000 articles. 
Some papers start to provide reviews~\cite{NatureMedecine20,harapan2020coronavirus,Huang2020.04.14.20065771} which are indeed very useful. This paper aims at providing a systematic methodology to mine such a large publication set while giving some specific focuses on topics of interest.

The rest of this paper is organized as follows: Section~\ref{sec:RelatedWork} presents some related works; Section~\ref{sec:Framework} describes the analysis framework including the data description and the methodology for analysis that was followed;  Section~\ref{sec:Method} describes the methodology we developed to get insights oo a specific sub-topic or on latest research. Section~\ref{sec:Origin} focuses on a deeper analysis on the origins of COVID-19.  Section~\ref{sec:Latest} focuses on the analysis of the terminology related to the latest research. Finally Section~\ref{sec:Conclusion} concludes this paper.

\section{Related work}
\label{sec:RelatedWork}

\textbf{Publication analysis on COVID-19.}
Some related work reports either studies on various topics or focuses on one or two topics. For example, the report from Nature Medecine \cite{NatureMedecine20} presents a focus on clinical trials and human studies, Preclinical studies, and Epidemiology. Another distinction on related work is what level as automatic assistance they benefit from. In most of the cases, it is difficult to know that level as it is not necessarily depicted. For example,  \cite{NatureMedecine20} seems to be a manual analysis of a few papers on the topic since the references are few.
Harapan \textit{et al.} \cite{harapan2020coronavirus} presents a literature review on Coronavirus disease 2019 (COVID-19). This study cites 74 references and presents a comprehensive state of the art on different sub-topics such as COVID-19 transmission, risk factors, diagnosis or treatments. 
Huang \textit{et al.} \cite{Huang2020.04.14.20065771} reviewed 1,281 abstracts from which they identified 322 manuscripts relevant to 5 areas of interest for their study. The five topics they chose are as follows: antibody kinetics, correlates of protection,  immunopathogenesis, antigenic diversity and cross-reactivity, and population seroprevalence. Dousset \textit{et al.} \cite{Dousset20} presents a short analysis of a larger dataset which size is similar to the one that we are using in this study. However their study does not go in deep in the publication content but rather analyze the collaborations between researchers and countries. They do not analyze any specific topic.

Different from the previsous analysis, this paper combines (a) the analysis of a large data set of about 25,000 references and (b) highly assisted analysis. While the methodology could be apply to various COVID-19 subtopics, we focus here on two main topics: the origin of the virus and the use of other disease treatment.

\textbf{Mining scientific publications.}
Most of the work in automatic publication analysis is related to scientometrics~\cite{osabe2018introductory} which as been defined as ``quantitative study of science, communication in science, and science policy''~\cite{hess1997science}. In this paper, publication mining is not strictly a question of scientometrics, rather the objective is to build knowledge from a large set of publications.

Shibata et al.  \cite{shibata2009detecting} uses citation network analysis in order to detect emerging research. Small et al. also use direct citation and co-citation analysis in order to identify emerging topics~\ref{small2014identifying}. Unfortunately, this information is not necessarily provided in digital database for large quantities of documents. 
Buscaldi et al. mine scholarly publications to build scientific knowledge graphs~\cite{buscaldi2019mining}. As in our approach, the authors use existing natural language processing and mining tools in their approach; however, they focus on the textual parts of the publications. 
Ronzano and Saggion developped a platform to automatically extract and enrich structural a,d semantic azspects of scientific publications. Their approach also focuses on the textual content and their applications are related to rhetorical sentence classification and extractive text summarization~\cite{ronzano2016knowledge}. 
Mothe et al. present a platform to mine scientific publications. In their paper, they focus on detecting the main collaborations, focusing on the geographical structure of a domain~\cite{mothe2006combining}.

\section{Analysis Framework}
\label{sec:Framework}

\textbf{COVID-19 scientific publication set.}
Recently, publishers have released the COVID-19 Open Research Dataset. It is available at \url{https://pages.semanticscholar.org/coronavirus-research}. This data set consists of multiple files. 

Among them, the Metadata file (60Mb) is a CSV file corresponding to 44,270   research articles with links to PubMed, Microsoft Academic and the WHO COVID-19 database of publications. The fields of the structure of the records are as follows: title, doi, abstract, date of publication, authors, journal, as well as internal document ids (PMC ID, PUBMED ID, Microsoft Academic Paper ID, WHO ID) and information whether the full text is available or not.
While the  meta file is a rich source of information, other pieces of information that are missing in that data file can be very useful such as the affiliation of the authors for example. For this reason we also considered a more complete set regarding the attributes that are provided. 

We chose to focus on the documents from PubMed only \footnote{\url{https://www.ncbi.nlm.nih.gov/pubmed}}, which is known to be realable source. It does not contains all the 44k scientific papers from the COVID-19 Open Research Dataset but about 24,000 papers.  
The query used to query the PubMed collection (\url{https://www.ncbi.nlm.nih.gov/pubmed}) was the same as the one used in the submentioned data set: \\
\textit{"COVID-19" OR "covid19" OR "Coronavirus" OR "Corona virus" OR "2019-nCoV" OR "SARS-CoV" OR "MERS-CoV" OR “Severe Acute Respiratory Syndrome” OR “Middle East Respiratory Syndrome”}
\\

\textbf{A few elements of the document collection.}
A few statistics are presented in Table~\ref{tab:Statistics} where the top 7 authors (the ones that authored the larger number of publications in the analyzed collection} are listed in Table~\ref{tab:TopAuthors}. In average, there is $5.6$ authors per publications. Not all the publications come with an abstract (72\% have an abstract for a total of 17,116). 
Table~\ref{tab:10Most} lists the most frequent venues where the scientific papers have been published.
\begin{table}[!ht]
\center 
\caption{A few statistics on the data collection.}
\label{tab:Statistics}
\center
\begin{tabular}{llll}
\bf  &  & Occur. at   &  at least \\
Label&Number  of & least twice & 10 times\\
\hline
Publication & 23,784  \\
Abstract & 17,116\\
Author&75,147\footnote{An author may appear in the collection with different spellings; for example there are 6 spellings for R.A. BARIC and two for L. ENJUANES }&9,211 & 1,145\\
Venues&3,120&1,832&431\\
Year&60\footnote{While the collected papers range from 1952 to 2020, 15 years have more than 500 publications.}&54&48\\
\end{tabular}
\end{table}

\begin{table}[!ht]
\center
\caption{Authors (full author names) with more than 100 publications in the collection.}
\label{tab:TopAuthors}
\begin{tabular}{l}
\bf Authors   \\ 
\small {
YUEN, KWOK-YUNG (Univ. of Hong-Kong)\\
PERLMAN, STANLEY (Univ. of Iowa)\\
DROSTEN, CHRISTIAN (Berlin Institute of Virology, Germany)\\
BARIC, RALPH S. (Univ. of North Carolina, USA) \\
MEMISH, ZIAD A (Alfaisal University, Saudi Arabia)\\
JIANG, SHIBO (New York Blood Center, USA)\\
ENJUANES, LUIS (Campus Univ. Autónoma de Madrid, Spain)\\}
\hline
\end{tabular}
\end{table}

\begin{table}[!ht]
\center
\caption{Most frequent publication venues  in the collection.}
\label{tab:10Most}
\begin{tabular}{ll}
\bf Venues  \\ 
 \small{JOURNAL OF VIROLOGY}\\
 \small{ADVANCES IN EXPERIMENTAL MEDICINE AND BIOLOGY}\\
 \small{VIROLOGY}\\
  \small{EMERGING INFECTIOUS DISEASES}\\
  \small{BMJ (CLINICAL RESEARCH ED.)}\\
  \small{JOURNAL OF MEDICAL VIROLOGY}\\
  \small{THE JOURNAL OF GENERAL VIROLOGY}\\
\hline
\end{tabular}
\end{table}


\\

\textbf{Topics of interest}.
While researchers have their topic of interest driven by their funding, project and research, topic of interest also come from the civil society on COVID-19. 

It is worth to mention that NIST/TREC has formed a joint effort called TREC-COVID\footnote{\url{https://ir.nist.gov/covidSubmit/}}. Like the other TREC tracks\footnote{TREC: Text Retrieval Conference is  co-sponsored by the National Institute of Standards and Technology (NIST) and U.S. Department of Defense supports research  for large-scale evaluation of text retrieval methodologies~\url{trec.nist.gov}}, TREC-COVID aims at gathering research teams in information retrieval to evaluate search engines on specific tasks. Mid April 2020, TREC-COVID has release a set of 30 topics of interest. A topic of interest is for example the ``Coronavirus origin'' or ``early symptoms''.

These topics also echo the most popular questions web searchers are interested on. Google trends mentions "Where  did coronavirius start?" and "How to know if you have coronarius" among the most asked questions.

Finally, these topics also echo the ones that have been considered in some related work papers \cite{harapan2020coronavirus,NatureMedecine20} that generally target the COVID-19 origins, its symptoms, spreading, risk factors and treatments.

\section{Methodology}
\label{sec:Method}

\textbf{Overview.}
The information we use is the data as collected. While automatic analysis helps in handling large quantities of publications, the conclusions drawn have to be  handle with  caution because there is no manual analysis of the content and no checking. Moreover, we did not solved content anomalies such as variants of entities spelling (e.g. author names). There are also missing values that we did not consider either and not resolved (See Table~\ref{tab:Statistics}).

The methods we use are usual data mining tools like frequency, graph-based visualization, factorial analysis. We consider crossing meta-data with content-based information from free texts as detailed later on.

The analysis considers both meta-data, such as the source, publication date and author fields and free text data such as the title and  abstract fields.
With regard to the titles and abstracts, we consider both single words as well as phrases that we automatically extracted. 
\\

\textbf{Keyphrase extraction}.
Different methods have been developed to extract key-phrases from free texts~\cite{hasan2014automatic,beliga2014keyword}; among the most popular are graph-based extraction\cite{boudin2013comparison,boudin2018unsupervised,mothe2018automatic}, co-occurrence-based methods~\cite{kaur2010effective} and more recently embeddings~\cite{10.1007/978-3-030-45442-5_41}. 

In our approach, we use a n-gram word extraction where we skip stop words. More precisely, we extract the most frequent n-grams after stop word removal, but without stemming to keep more precise semantics. We also consider an initial lexicon  from composed terms (as written by the authors e.g. ``anti-malagia'' or ``animal-origin'') as initial phrases that are enriched by the n-gram extracted ones. \\

\\

\textbf{Graph-based visualization.}
Graphs are among visualization tools the most used in the literature, as linking concepts
or objects is the most common mining technique~\cite{mothe2006combining}.
Graph-based visualizations are widely used to visualize bibliometric networks~\cite{vanEck2014,van2017citation}. 

In this paper, we mainly use bipartite graphs.
A bipartite graph is a graph whose vertices (nodes) can be divided into two disjoint and independent sets and where edges connect a node of each type. A bipartite graph does not contain any odd-length cycle (Wikipedia). This type of representation is also widely used for document analysis and visualizations \cite{crossno2011topicview,alsallakh2014visualizing}.
In this paper, bipartite graphs are used to visualize the results of crossing meta-data and keyphrases extracted from publications.\\

\\ 

\textbf{Process.}
In this paper, we developed two different processes: the first one can be applied to focus on any specific sub-topic of interest, this is the process we applied to the "Origin" of the virus sub-topic (See Section~\ref{sec:Origin}). The second process is related to the latest research and we apply it to detect some topic clusters (See Section~\ref{sec:Latest}).

\subsubsection*{Getting insights on a specific topic}

This process consists into three steps:
\begin{itemize}
    \item Select the keyphrases related to the topic of interest. It can be computer-assisted by considering strings of characters (e.g. "ORIGIN" is a relevant character string to extract many relevant key-phrases such as "BAT-ORIGIN", "HUMAN-ORIGIN",...) ; stems of the topic word(s) is a good start. The automatically obtain list should then be manually checked in order to remove non-relevant terms (e.g. "ORIGINALITY")
    \item Build a bipartite graph where vertices are  key-phrases in the one hand and publication identifiers in the other hand. This first representation provides a quick overview of the use of the terms: are the terms shared among publications or are they rather partitioned (each publication is more focusing on one of the aspects). It is also a mean to directly go to the associated publications;
    \item For one specific term, build the bipartite graph where the other vertices are authors names. This graph can be weighted by the number of publications of each author that mention the terms. This graph shows the most "important" authors related to that term (who are likely to be specialists). Such a graph can also be built considering several related terms at the same time when terms are non independents in publications (cf previous step). The later graph makes it possible to highlight the authors that tackles several aspects of the topic.
\end{itemize}{}
Other meta data could  be also crossed to include additional steps in that process (e.g. considering the authors' affiliation or affiliation countries). This step was not included in this paper. 

\subsubsection*{Getting the latest research topic of interest}

This process also consists into three steps:
\begin{itemize}
    \item Extract the latest terminology: these are the keyphrases that occurs in the latest year(s) but not before. They are obtained by crossing the keyphrases and the year of the publications the keyphrases occur in. We consider also the occurrence frequency when selecting them;
    \item Extract highly connected graphs (``communities'' of keyphrases). This step results in sub-topics of interest consisting of terms of various nature but often used together in recent publications;
    \item Build  bipartite graphs where vertices of the first type are the terms from one community and vertices of the second type are publication identifiers to identify the relevant publications with regard to the group of terms.
\end{itemize}{}

This process is used to get the latest research topics and associated publications.

\section{Origins }
\label{sec:Origin}

\begin{figure}[!h]
    \centering
  \hspace*{5mm}  
  \subfigure[a]{\includegraphics[width=0.30\textwidth]{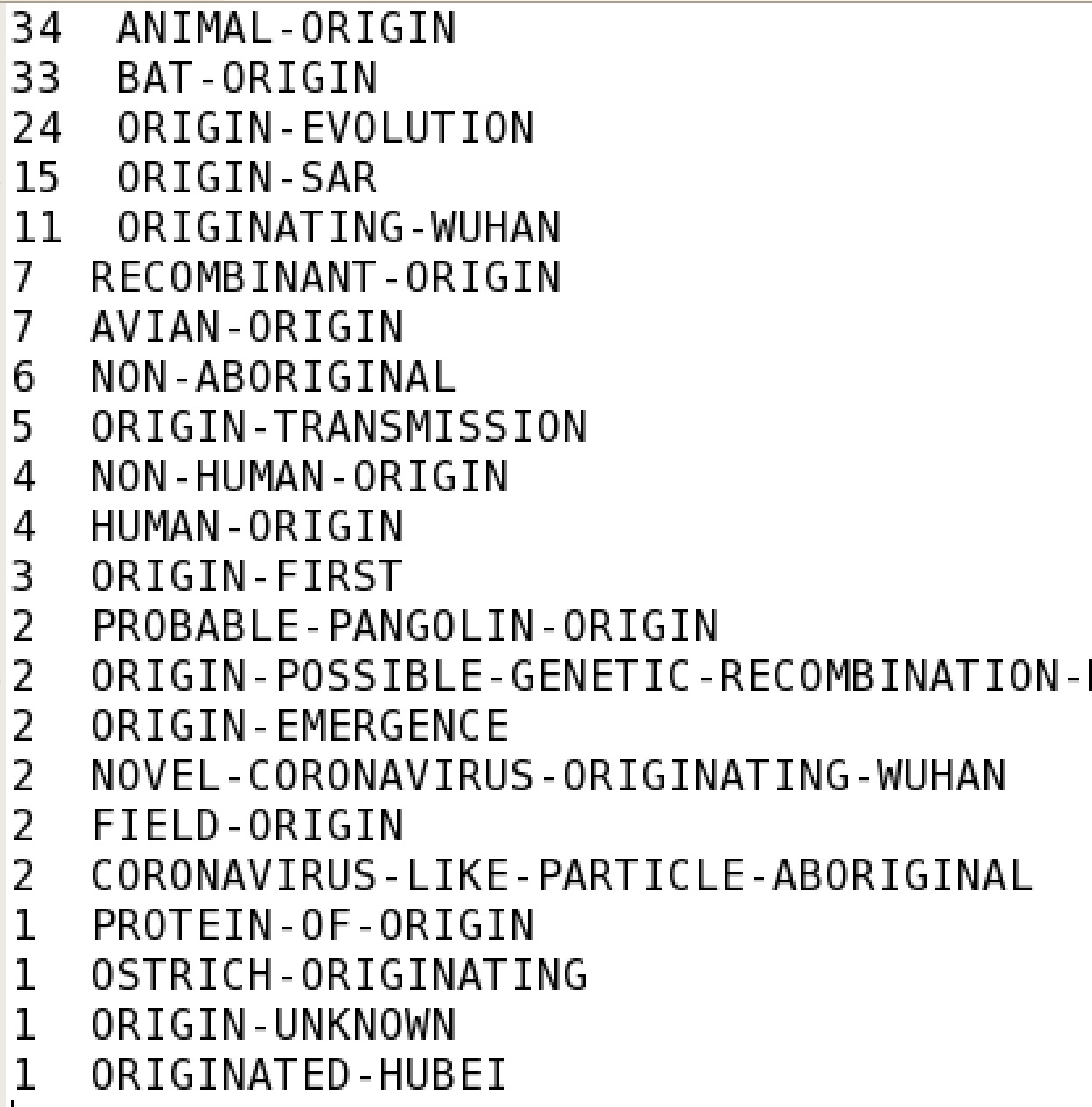}
  \subfigure[b]{\includegraphics[width=0.15\textwidth]{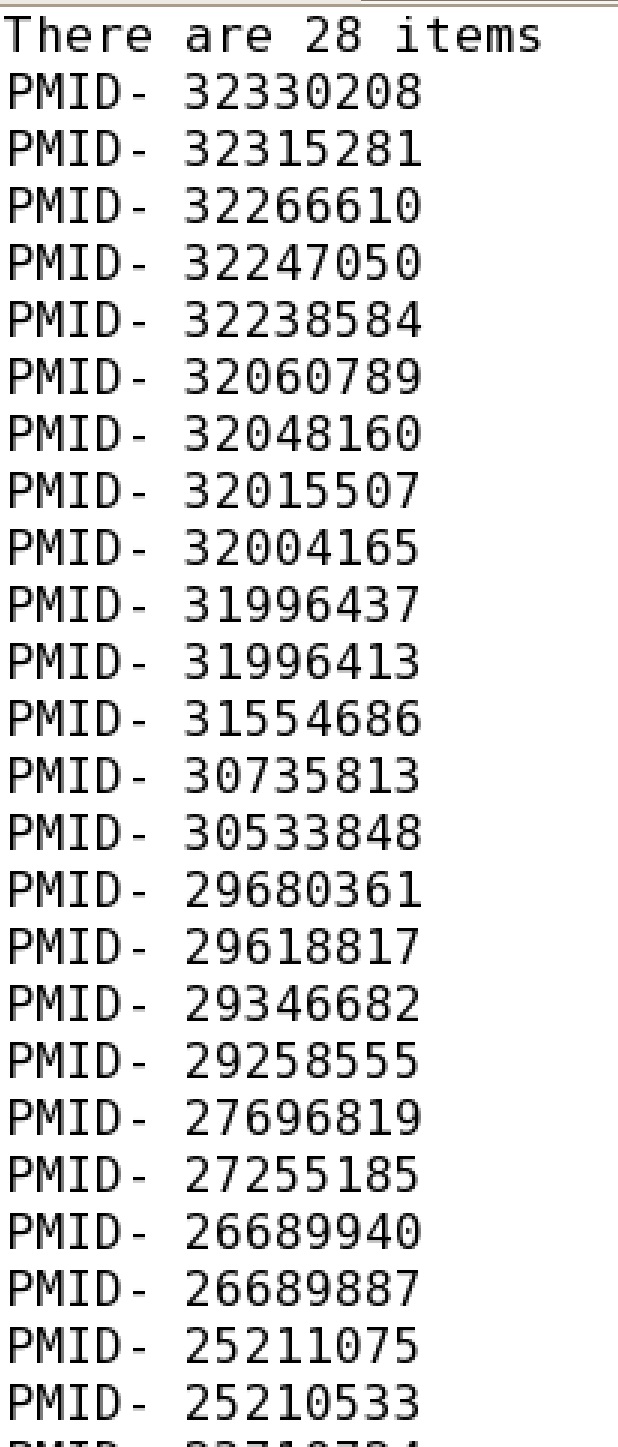}\\
    \subfigure[c]{\includegraphics[width=0.65\textwidth]{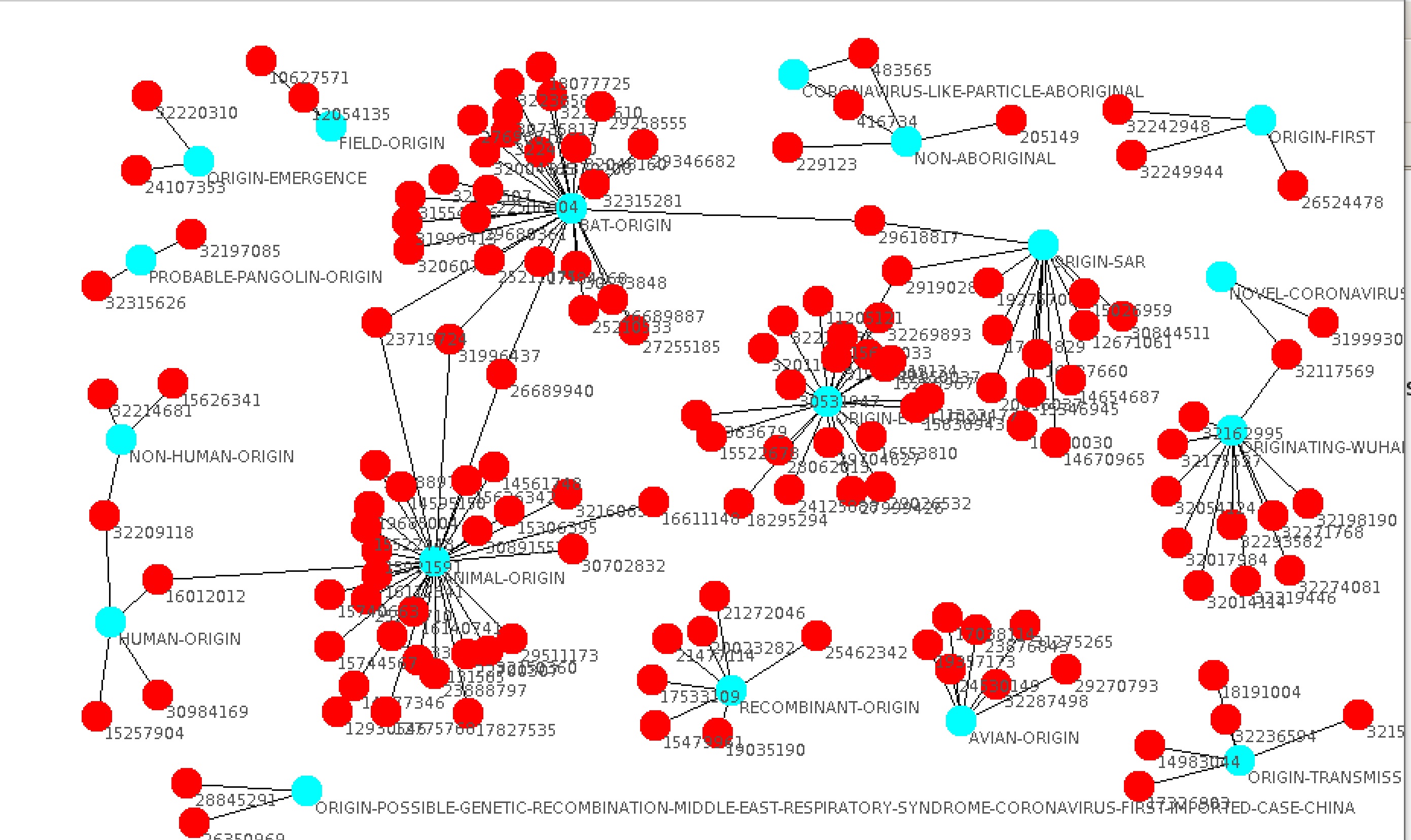}} 
     \caption{(a) Terms related to the ORIGINs of the virus (b) Publication PMID that mention BAT-ORIGIN (c) Graph of the ORIGIN terms and associated PUB-ID  }
    \label{fig:Origin1}
\end{figure}

The  topic of interest we tackle in this section is related to the COVID-19 origins and uses the first process as described in Section~\ref{sec:Method}.

Figure~\ref{fig:Origin1}(a) displays the terms related to the ORIGINs of the virus (extracted from the publications titles and abstracts when available) as well as their frequency. For example, 33 publications study BAT-ORIGIN and 7 AVIAN-ORIGIN. We also used these terms to extract the associated publications. These terms are used in the graph from Figure~\ref{fig:Origin1}(c) as blue nodes. Figure~\ref{fig:Origin1}(b) displays the PMID of the publications that mention the "BAT-ORIGIN".

Figure~\ref{fig:Origin1}(c) is a bipartite graph where blue nodes are  the ORIGIN terms while the red nodes are the publication identifiers. This graph is not fully linked since some publications mention one of these "ORIGIN" terms only. From this graph, we can also identify the publications that mentions several of the ORIGIN terms as PMID- 26689940~ \cite{hu2015bat}, 31996437~\cite{wan2020receptor} and 23719724~\cite{lorusso2019discrepancies} which  mention both "ANIMAL-ORIGIN" and "BAT-ORIGIN". Another example is PMID- 16012012~\cite{lai2005low} which mention both "HUMAN-ORIGIN" and "ANIMAL-ORIGIN" (See Figure~\ref{fig:Origin1}(c)).

\begin{figure}[!h]
    \centering
  \hspace*{5mm}      \subfigure[a]{\includegraphics[width=0.60\textwidth]{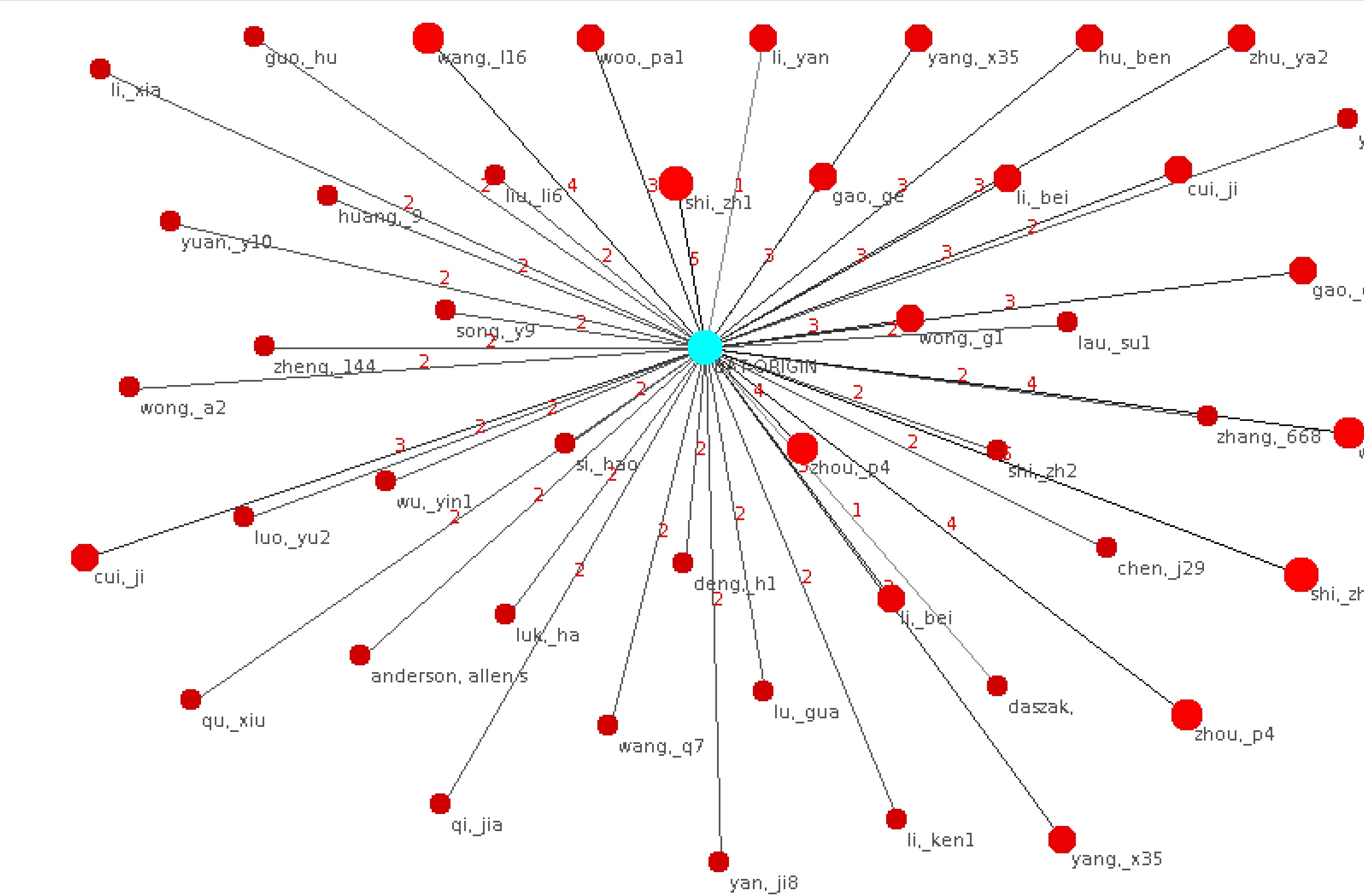} \\
    \subfigure[b]{\includegraphics[width=0.80\textwidth]{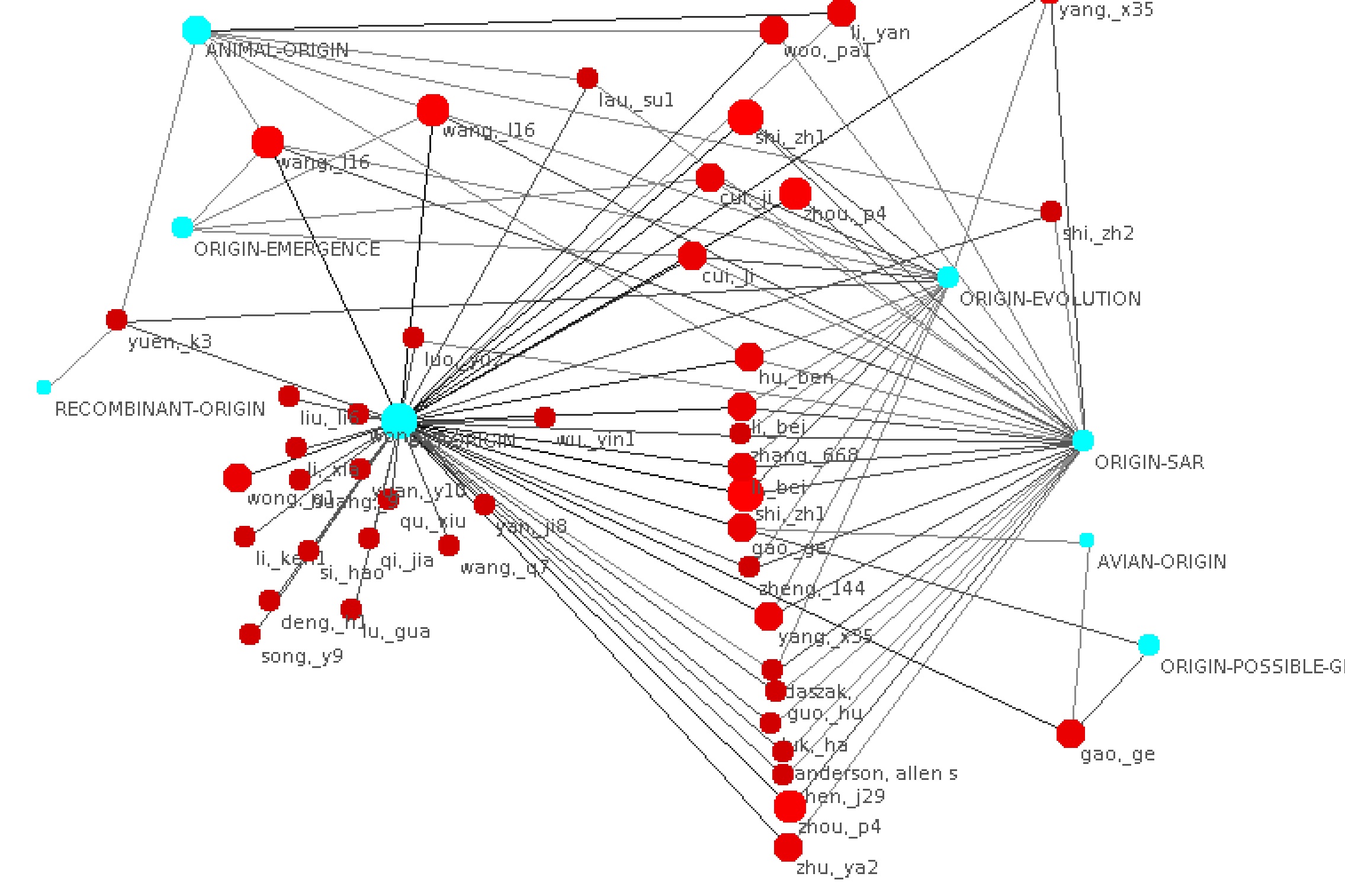}}
    \caption{Authors associated to (a)  BAT-ORIGIN term (b) ORIGIN terms from the one listed in Figure~\ref{fig:Origin1}  }
    \label{fig:Origin2}
\end{figure}

With regard to the publications associated with BAT-ORIGIN terms in  Figure~\ref{fig:Origin1}(b), we can mention Ren et al. \cite{ren2020identification} (PMID- 32004165) who report a novel bat-origin CoV causing severe and fatal pneumonia in humans. The identified virus is phylogenetically closest to a bat SARS-like CoV.  PMID-31996413 \cite{li2020discovery} employs a capture-based NGS approach for virus discovery. Since (SARS)-COV and (MERS)-CoV both originated from bats, active surveillance is recommended in this paper. Yang et al. \cite{yang2019broad} (PMID- 31554686) study potential cross-species transmissibility of SADS-CoV.  PMID- 30533848 \cite{li2018complete} studies the full-length genome sequence of a novel swine acute diarrhea syndrome     coronavirus (SADS-CoV), CH/FJWT/2018 which is closely related to CN/GDWT/2017. A link is also made with bat-origin SADS-related coronaviruses. Wang et al. (PMID- 29680361 \cite{wang2018bat}) report cross-species transmission due to a large number of mutations on the receptor-binding; here a novel bat-origin coronaviruses found in pigs is considered. Considering the the publications that mention BAT-ORIGIN of COVID, the oldest paper was published in 2006~\cite{feng2006baculovirus}, 11 were published in 2020, 2 in 2019 and 4 in 2018.

We also had a look to the associated authors (See Figure~\ref{fig:Origin2}(a)). In this figure, the only blue node is BAT-ORIGIN term, while the red nodes are authors of publications that mention this term. The value on the link indicates the number of publications a researcher authored that mention BAT-ORIGIN. Figure~\ref{fig:Origin2}(b) shows, for those authors that mention BAT-ORIGIN, the other ORIGIN terms also mentioned by them. The thickness of the link  is an indication of the number of publications as well as the size of the nodes.


\section{Latest research}
\label{sec:Latest}
While in the previous section we did not consider the year of publication, in this section, we focus on the latest research and the associated terminology.
We consider the publications that are published in the 30 last years only (1991 to 2020). We then keep the only phrases or automatically extracted keywords from the titles and abstracts (see Section~\ref{sec:Framework}) that mainly occurs in 2020. More precisely, to be kept, a keyphrase has to occur at least 80\% times in 2020. These keyphrases (there are about 1,500) are thus the keyphrases of current interest. We then built a graph where nodes ( keyphrases) are linked when co-occuring in a publication. This is a weighted graph (the more the two keyphrases co-occur, the higher the weight). We then extracted clusters of terms that are closely related as communities (nodes that are highly inter-connected together and weakly connected with other nodes) considering different focuses.

    


\begin{figure}
    \centering
  \hspace*{5mm}      \subfigure[a]{\includegraphics[width=0.45\textwidth]{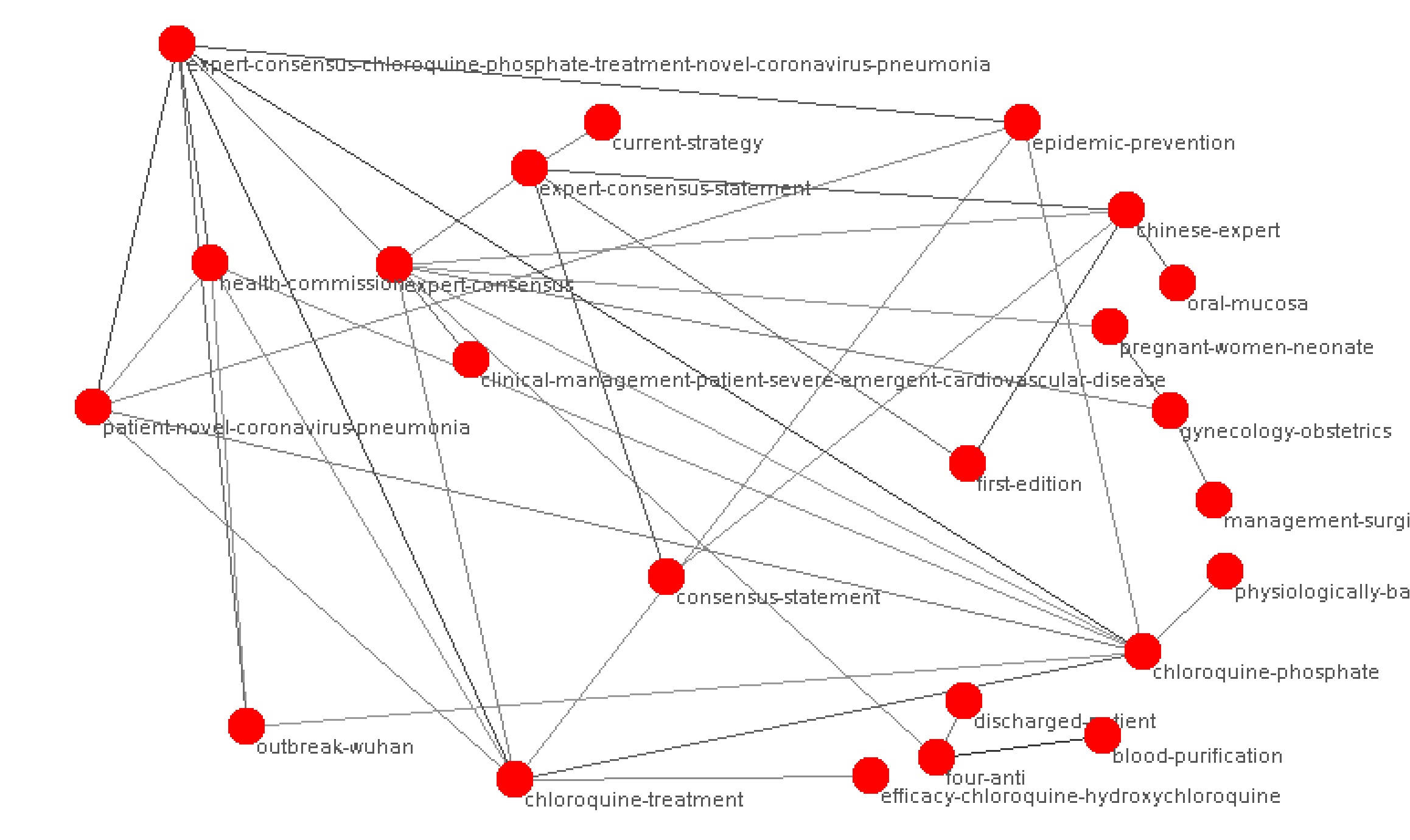} 
    \subfigure[b]{\includegraphics[width=0.45\textwidth]{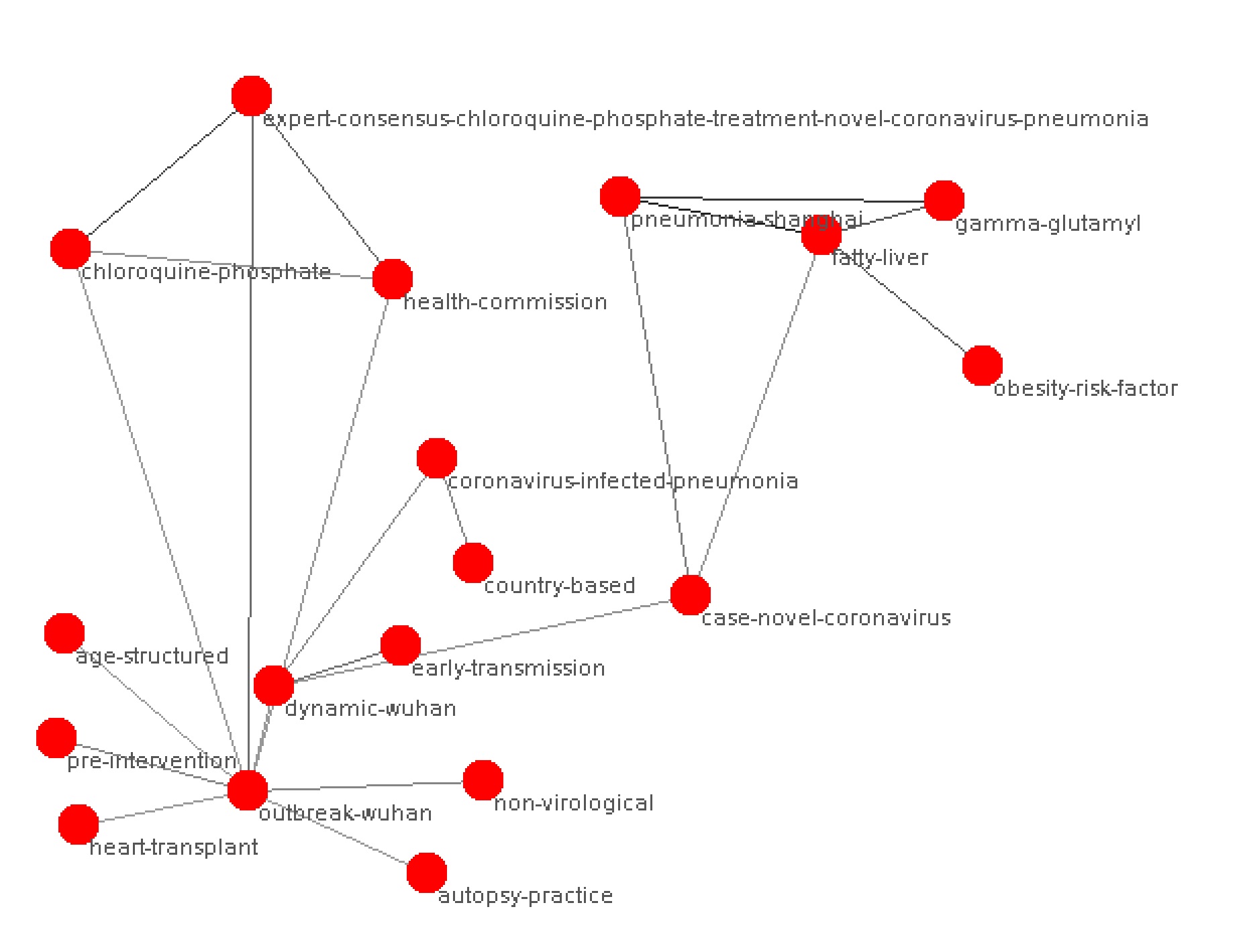}}\\
    \caption{(a) Chloroquine-treatment term network (b) Obesity Risk Factor network.}
    \label{fig:Latest1}
\end{figure}

The example in Figure~\ref{fig:Latest1}(a) focuses on the Chloroquine-treatment while Figure~\ref{fig:Latest1}(b) is related to Obesity risk factor. From these figures, we can also identify the related terms. 

\begin{figure}
    \centering
  \hspace*{5mm}           \includegraphics[width=0.90\textwidth]{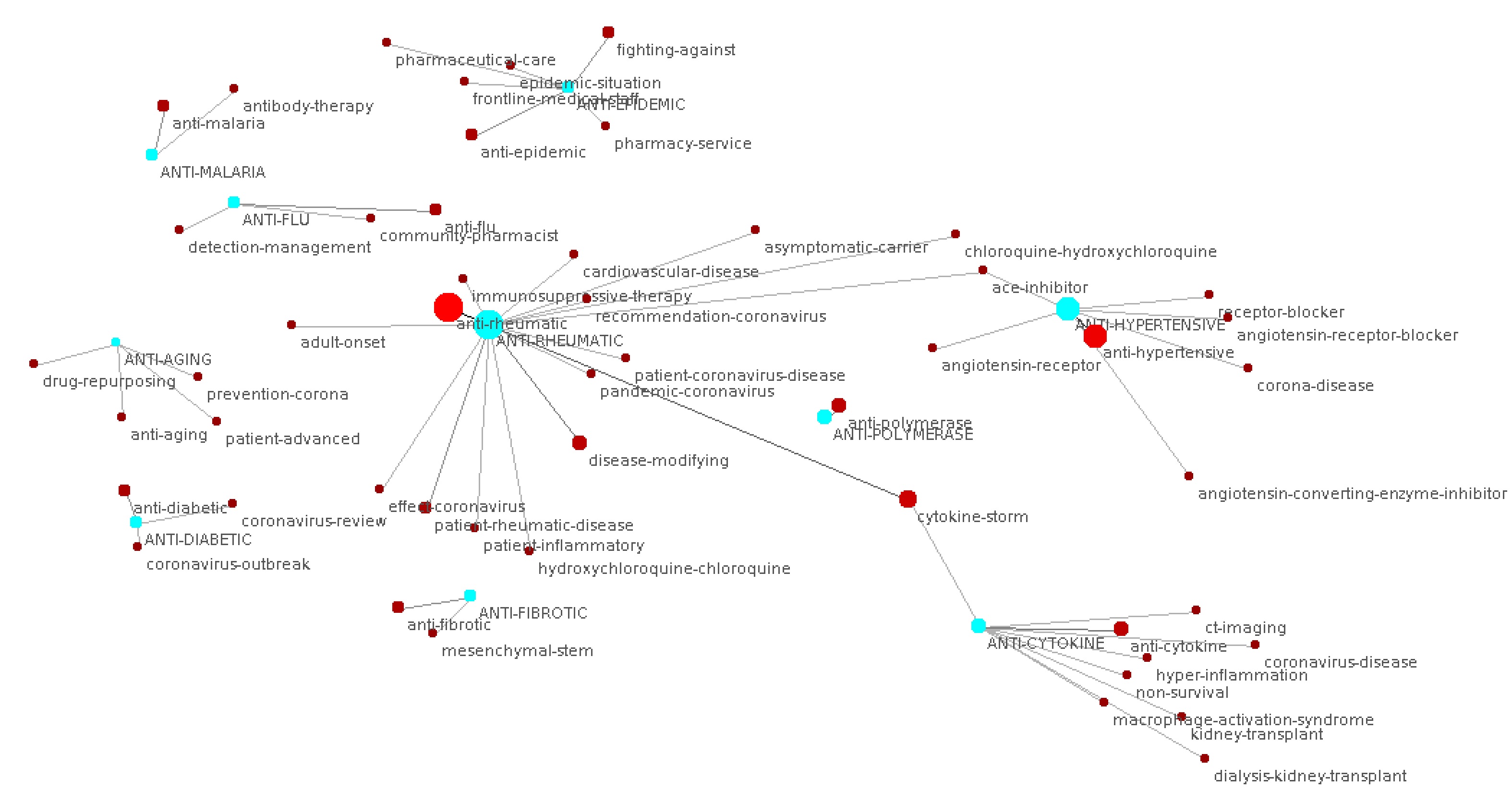}
    \caption{ Various "ANTI-" uses.}
    \label{fig:Latest3}
\end{figure}

\begin{figure}[!]
    \centering
  \hspace*{5mm}      
  \includegraphics[width=0.7\textwidth]{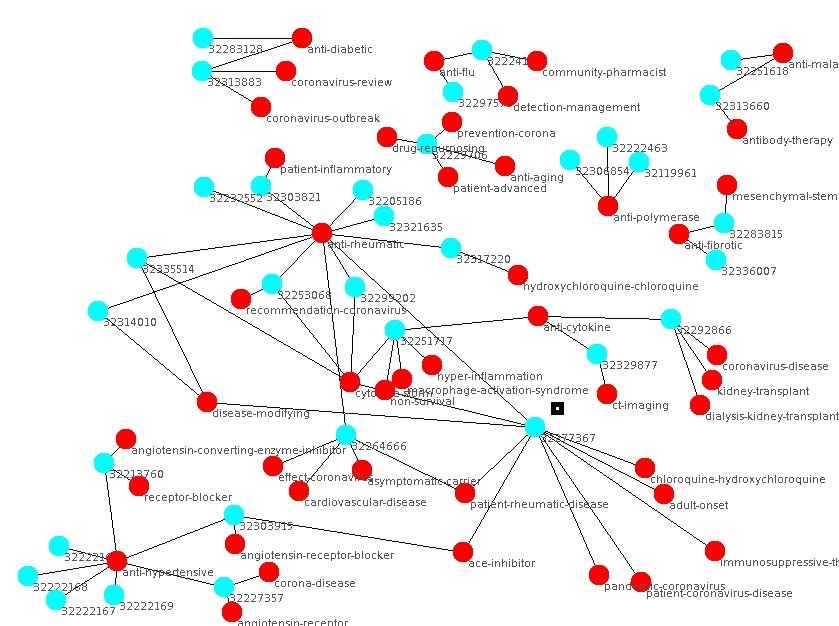}
   
    \caption{PMID of the publications that mention an "ANTI" terms and the other related terms.}
    \label{fig:Latest2}
\end{figure}

COVID-19 is an infectious disease caused by SARS-CoV-2 and several papers investigate the use of existing anti-viral treatments. For examples, some of the publications mention "anti-" (e.g. anti-flu, anti-malaria, ..). We thus had a closer look to the network related to treatment used for other diseases considering various "anti-" terms as shown in Figure~\ref{fig:Latest3}. 
As we did in Section~\ref{sec:Origin} and illustrated in Figure~\ref{fig:Origin1}(c), we look at the publications associated with these terms. We found 33 publications directly related to these terms. The PMID of these publications as well as related terms are presented in  Figure~\ref{fig:Latest2}.

From this Figure~\ref{fig:Latest2}, we can see that some of the papers are more connected than others; suggesting they potentially mention more topics of interest. For example,  PMID 32277367 \cite{misra2020rheumatologists}, which is marked up with a black square on the right-bottom part of Figure~\ref{fig:Latest2}, was published on the 10th of April, in J Neurooncol.  and results from a collaborative work of India, Russia, and UK and concludes that ``it may be cautiously recommended to continue glucocorticoids and other disease-modifying antirheumatic       drugs (DMARDs) in patients receiving these therapies, with discontinuation of DMARDs during infections as per standard practice''.  This paper also mentions various evidence that suggest potential benefits of various drugs. Reversely, we can see in the same figure that some keywords are also more shared than others. As an example, anti-rheumatic is mentioned in 9 of the displayed publications. 

Among these publications, we can quote Perricone \cite{perricone2020anti} (PMID- 32317220) discusses the anti-viral aspect of immunosuppressants for searching for a potential treatment for SARS-CoV-2 infection. Lehrer et al. \cite{lehrer2020inhaled} (PMID- 32313883) discusses the effects of biguanides on influenza and coronavirus. Kumar et al. (PMID- 32313660) study the antibody therapy as an immediate strategy for emergency prophylaxis and SARS-CoV-2 therapy~\cite{Kumar20}. Song et al. (PMID- 32314010) reports a case of COVID-19 pneumonia on a 61-year-old female rheumatoid arthritis; she was treated with antiviral agents      (lopinavir/ritonavir), and treatment with cDMARDs was discontinued except hydroxychloroquine. Her symptoms gradually improved and three weeks later, real-time PCR for COVID-19 showed negative conversion~\cite{song2020coronavirus}.



\section{Conclusion}
\label{sec:Conclusion}
This paper presents two analytical processes in order to mine scientific papers that are illustrated on COVID-19 scientific publications. The results are knowledge graphs of various natures that helps getting insights on specific subtopics or recent research topics. 
While scientific papers are reliable sources of knowledge, on COVID-19, other sources of information such as the World Health Organization reports would also worth being analysed using the same type of methodology.



\bibliography{MotheKes20}
\bibliographystyle{elsarticle-harv}

\end{document}